\documentclass[11pt]{article}
\usepackage[a4paper,margin=2.4cm]{geometry}
\usepackage[no-math]{fontspec}
\usepackage{amsmath,amssymb}
\usepackage[version=4]{mhchem}
\usepackage{graphicx,booktabs,xcolor,setspace,lineno,url}
\usepackage[super,sort&compress,comma]{natbib}
\usepackage{caption}
\DeclareCaptionLabelFormat{nature}{\textbf{#1 #2 |}}
\usepackage[hidelinks]{hyperref}
\begin{document}
\begin{center}
{\Large\bfseries U-HamGNN: a universal model for magnetic Hamiltonians conditioned on spin and Hubbard U\par}
\vspace{10pt}
{\normalsize Hongyu~Yu\textsuperscript{1}, Yinze~Hu\textsuperscript{1}, Yizhou~Chen\textsuperscript{1}, Yang~Cheng\textsuperscript{1}, Yingwei~Chen\textsuperscript{1,2}, Xingao~Gong\textsuperscript{1}, Yang~Zhong\textsuperscript{1,*}, Hongjun~Xiang\textsuperscript{1,*}\par}
\vspace{6pt}
{\small\textsuperscript{1}Key Laboratory of Computational Physical Sciences (Ministry of Education), Institute of Computational Physical Sciences, State Key Laboratory of Surface Physics, Department of Physics, Fudan University, Shanghai, China\par}
{\small\textsuperscript{2}Research Laboratory for Quantum Materials, Department of Physics and Materials, The Hong Kong Polytechnic University, Hong Kong SAR, China\par}
{\small\textsuperscript{*}Corresponding authors: \texttt{yzhong@fudan.edu.cn}, \texttt{hxiang@fudan.edu.cn}\par}
\end{center}
\vspace{6pt}

\section*{Abstract}
The electronic structure of a magnetic crystal depends jointly on its spin configuration and the on-site Hubbard correction U, requiring a separate DFT+U calculation for each combination. We introduce U-HamGNN, an equivariant network that takes these two inputs directly and predicts spin-resolved Hamiltonians across diverse crystal chemistries. A spectral-hybrid optimizer that weights each update direction by how consistently the gradient points along it reaches 8\% lower Hamiltonian error than Muon and 26\% lower than AdamW in a ten-epoch benchmark. The predicted bands reproduce U-dependent gap opening and altermagnetic spin splitting within the trained window and beyond it; exchange couplings and magnetic-order calls derived from the Hamiltonian agree with DFT+U on a calibration set. This universal model provides a new tool for magnetic research.

\section*{Introduction}

Magnetic materials carry the spin degrees of freedom behind permanent magnets, magnetic memory and spintronic devices. Altermagnets and magnetic semiconductors, recognized more recently, add spin-resolved electronic structures that these devices can use\cite{jungwirth2016,smejkal2022landscape,li2012bms}. Whether a compound is an altermagnet, a half-metal or a magnetic semiconductor, and how strongly its moments couple, is decided by comparing several collinear magnetic configurations of the same crystal. The calculation of each configuration also depends on the on-site Hubbard correction U that density functional theory plus U (DFT+U) applies to partially filled d and f shells. Its value shifts gaps and moments and can reorder the magnetic states\cite{anisimov1991,dudarev1998,cococcioni2005,himmetoglu2014}, and since experiment does not fix it, U has to be scanned. The electronic structure of one magnetic material is therefore a function of its magnetic state and of U. A high-throughput survey needs one self-consistent calculation per structure, configuration and U value, and magnetic calculations are harder to converge than non-magnetic ones\cite{horton2019}.

Machine learning has lowered the cost of first-principles calculations by learning their outputs directly\cite{butler2018}. Predicting the electronic Hamiltonian in an atomic-orbital basis is the most reusable choice: bands, spin splittings and exchange couplings then follow from it without a self-consistent cycle. Neural networks that respect the rotational symmetry of the orbitals predict such Hamiltonians within one material family with errors of a few meV\cite{schutt2019schnorb,unke2021phisnet,li2022deeph,gong2023deephe3,zhong2023hamgnn,yu2023qhnet}. Universal models trained on tens of thousands of crystals now cover most of the periodic table and run without retraining on databases of predicted crystals such as GNoME\cite{zhong2024universal,wang2024universal,merchant2023}.

Magnetism has entered these models only in part. xDeepH\cite{li2023xdeeph} and HamGNN++\cite{zhong2023hamgnnpp} learn how the Hamiltonian of one magnetic system depends on its spin configuration, and magnetic neural network potentials learn energies as functions of the moments\cite{yu2022spinham,yu2024spingnn,yu2024spingnnpp,eckhoff2021mhdnnp,novikov2022mmtp,rinaldi2024magace,yang2024deepspin}. Each is trained on one material family with a single exchange-correlation setting, in which U is fixed. The universal Hamiltonian models are non-magnetic; our spin-orbit-coupled Uni-HamGNN extended universality to relativistic Hamiltonians and named magnetism as its next step\cite{zhong2026unihamgnn}. Universal interatomic potentials carry magnetic moments only as auxiliary labels\cite{deng2023chgnet}. The dependence of the Hamiltonian on the magnetic configuration and its dependence on U have each been learned for single systems, but no transferable model has learned both.

Whether one network trained across the periodic table can learn how the Hamiltonian of a magnetic crystal depends jointly on its collinear configuration and on U has not been tested. The test is whether the band gaps, spin splittings, exchange couplings and magnetic order derived from its predicted Hamiltonian agree with self-consistent DFT+U for the same structure, magnetic configuration and U. Agreement on these quantities would let a magnetic search screen with the network first and run DFT+U only on the candidates it ranks highest, instead of computing every combination.

U-HamGNN uses the equivariant architecture of Uni-HamGNN and is trained on magnetic DFT+U data in which most structures appear at several U values and in several magnetic states. The network therefore takes the magnetic state and U as inputs, whereas earlier models fixed at least one of them in the training data. Each atom's U enters through a Gaussian embedding that vanishes at U = 0, and the two spin channels are built by adding learned exchange-like couplings, weighted by the input spins, to a non-self-consistent reference Hamiltonian. At batch size one and more than a hundred thousand magnetic training examples, training is limited by the dense matrices that generate the radial weights: they hold most of the parameters, and their gradient directions drift from step to step. We therefore developed a spectral-hybrid optimizer that weights each singular direction of the update by how consistently the gradient points along it, so that persistent directions grow and noisy ones are damped. We present the architecture and training first, then test the network in four steps of increasing scope: accuracy across the periodic table on test graphs; its response to U, by following one crystal across U against DFT+U; the exchange couplings and magnetic order derived from the predicted Hamiltonian, against DFT+U on a calibration set; and DFT+U adjudication of the GNoME altermagnet candidates that the model ranks highest. The model is collinear, has no spin-orbit term and is trained for U up to 8 eV.

\section*{Results}

\subsection*{Model architecture and training data}

U-HamGNN predicts the spin-resolved DFT+U Hamiltonian of a crystal in the OpenMX atomic-orbital basis\cite{ozaki2003,ozaki2004,han2006} from three inputs: the crystal graph (atoms as nodes, neighbour pairs within a cutoff as edges), a collinear spin direction $s_i \in \{+1, -1, 0\}$ on each site and the effective Hubbard interaction $U_i$ on each site (Fig. 1a). It uses the equivariant construction of HamGNN and Uni-HamGNN\cite{zhong2023hamgnn,zhong2026unihamgnn} on e3nn irreducible representations\cite{geiger2022e3nn,thomas2018tfn,batzner2022nequip} with OpenEquivariance kernels\cite{bharadwaj2025oeq}. U enters at the input layer: each atom carries a scalar attribute that joins its element embedding to a Gaussian expansion of its own U,

\begin{equation}
a_i = \big[\, e(Z_i)\, ;\; g(U_i) - g(0) \big], \qquad g_k(U) = \exp\!\big[-(U-\mu_k)^2 / 2\sigma_U^2\big], \quad \mu_k \in [0, 8]\ \mathrm{eV},
\end{equation}

with 32 evenly spaced centres $\mu_k$ and a width $\sigma_U = 8/31 \approx 0.26$ eV equal to their spacing, so an atom without a Hubbard correction keeps the plain element embedding. Three equivariant convolution layers with edge updates produce node and edge features, and two heads read orbital blocks from them through Clebsch–Gordan coupling (Fig. 1b). One head returns a spin-independent block $H^{\mathrm{nm}}$ for every site and bond. The other returns a coupling $J$ per pair of orbital shells and a learned weight matrix $W$; combined with the spins of the magnetic atoms $\mathcal{M}$, they give a collinear term that splits the two channels, the collinear form of the magnetic parameterization of our earlier model\cite{zhong2023hamgnnpp} (Methods),

\begin{equation}
H^{\sigma}_{X} = H^{0}_{X} + H^{\mathrm{nm}}_{X} + \sigma\,\Delta_{X}, \qquad \Delta_{X} = W_{X} \circ \sum_{a \in \mathcal{M}(X)} J_{X,a}\, s_a, \qquad \sigma = \pm 1,
\end{equation}

where $X$ is a site or a bond and $H^{0}$ is the non-self-consistent reference Hamiltonian that OpenMX builds from atomic densities. The network therefore learns the self-consistent DFT+U correction to a reference it is given. Because the spins and U are inputs, our model, with 23.65 million trainable parameters, can be queried for any collinear configuration and any U between 0 and 8 eV.

The training set holds 35,632 magnetic crystal structures, Materials Project and MAGNDATA-derived cells\cite{jain2013,gallego2016} covering 71 elements, computed with OpenMX DFT+U (Supplementary Fig. S6). Each structure at one U setting and in one magnetic state is one training graph, 154,546 in all. Most structures appear at two or three U settings (9,833 and 20,017 of the 35,632, against 5,782 at a single setting) and in several collinear magnetic states. The network thus sees the same geometry with different moments and U, and has to attribute the change in the Hamiltonian to those inputs. The model was trained on 128 NVIDIA H100 GPUs, and every result below uses one frozen checkpoint, Stage-2 epoch 78.

\subsection*{Training with the spectral-hybrid optimizer and test accuracy}

About 60\% of the parameters sit in the dense matrices that generate the radial weights, and at batch size one their gradients change direction from step to step. A coordinate-wise optimizer such as AdamW\cite{loshchilov2019adamw,kingma2015adam} scales each parameter separately and ignores the matrix structure of the update. Orthogonalized-momentum methods such as Muon\cite{liu2025muon,bernstein2024norm} equalize the singular values of the update but cannot tell a persistent direction from a noisy one. Our spectral-hybrid optimizer keeps a streaming rank-$r$ basis of the momentum, compares the gradient with a lagged estimate along each basis direction and reweights the update accordingly,

\begin{equation}
W_{t+1} = W_t - \eta\, c\, U_t\, \mathrm{diag}(w)\, V_t^{\mathsf T},
\end{equation}

where $U_t$ and $V_t$ come from one power iteration on the momentum, the weights $w$ grow where the two estimates agree, shrink where they disagree and equal one when reweighting gains nothing, and $c$ applies the root-mean-square scaling of Muon (Methods, Supplementary Fig. S3). The 48 radial-weight matrices take this update and the other 80 tensors take AdamW. We compared it with Muon and AdamW on 500 iron-containing graphs (Fe-500) with a 33.2-million-parameter configuration trained for ten epochs. It reached a validation Hamiltonian error of 23.5 meV, 8.4\% below Muon (25.7 meV) and 26\% below AdamW (31.7 meV) (Fig. 2a). In a smaller 14.2-million-parameter configuration with identical settings for every optimizer and three seeds, its full-rank variant also had the lowest error in every seed, with means of 27.3 meV against 27.7 meV for Muon and 31.0 meV for AdamW (Supplementary Fig. S3b). The production run used it in both training stages. The validation error fell from 2.78 to 2.18 meV in the Hamiltonian-only stage, and in the band-energy stage the band error fell from 348 to 96 meV, reaching its minimum at Stage-2 epoch 78 (Fig. 2b).

On the 1,577 test graphs our model reaches a mean absolute error of 1.80 meV over all 1.48 billion nonzero Hamiltonian matrix elements of their 4,073,125 orbital blocks (the Hamiltonian matrix of one atom or atom pair). Predicted and reference elements lie on the identity across 24 eV with a coefficient of determination of 0.99984, and this includes the small elements within 0.5 eV of zero that carry the hopping (Fig. 2c). Per graph, the error has a median of 1.67 meV, a mean of 2.05 meV and a 95th percentile of 4.30 meV (Fig. 2d). By element, it is 1.65 to 2.35 meV for nickel, manganese, iron and cobalt and between 1.45 and 6.35 meV for every element, highest for technetium and rhenium (Fig. 2e). On test graphs, eigenvalues at off-axis k points have a median error of 55 meV (Supplementary Fig. S5). The error changes little across the trained U window and between antiferromagnetic-like and ferromagnetic-like inputs, and it is highest at the 8 eV ceiling (Supplementary Fig. S4).

\subsection*{Band gaps and spin splitting across Hubbard U}

To test whether the network has learned how the Hamiltonian responds to U and to the magnetic state, we followed one crystal across U and compared its predicted bands with converged DFT+U at the same structure, magnetic state and U, on the same k path (Fig. 3). \ce{MnTe} is an altermagnet whose spin-split bands have been measured\cite{krempasky2024mnte}. At U = 0 the model and DFT+U both give a metal, with a band crossing the chemical potential. By U = 2 eV a gap has opened in both, and it widens with U, from 0.73 to 1.02 eV in the model against 0.72 to 1.10 eV in DFT+U between 2 and 8 eV. Bands within 1 eV of the chemical potential differ from DFT+U by 21 to 59 meV on average across the five U values (Fig. 3b; Supplementary Table S8). The model also reproduces the altermagnetic splitting. Along Γ–M–M$'$–Γ–A the two DFT+U spin channels are degenerate to 0.01 meV, and away from these lines they split. The largest same-band splitting within 2 eV of the chemical potential stays within 66 meV of DFT+U at every U, and both show a dip at 6 eV (830, 764, 725, 468 and 624 meV in the model against 864, 781, 711, 453 and 690 meV in DFT+U from U = 0 to 8 eV). The k-resolved splitting follows the DFT+U curve point by point (Fig. 3b, lower strips).

We followed a larger altermagnet candidate, the 32-atom cell of \ce{Ag3RuO4}, across six U values from 0 to 10 eV, beyond the 8 eV limit of the training data: its same-band splitting is 64 to 102 meV in the model against 66 to 124 meV in DFT+U, with mean band errors of 22 to 82 meV (Supplementary Fig. S1f). The model reproduces its gap of about 0.1 eV at U = 0, 2 and 8 eV, underestimates it at 4 eV (0.06 against 0.09 eV) and closes it at 6 eV, where DFT+U gives 0.12 eV (Fig. 3c). At 10 eV, outside the trained window, it still gives a gap of 0.08 eV and a splitting of 102 meV, against 0.11 eV and 124 meV in DFT+U. Beyond the trained window the model thus still predicts a gapped, spin-split state for this crystal, with a splitting 18\% below DFT+U. Band comparisons for further compounds are given in Supplementary Fig. S1.

\subsection*{Exchange couplings and magnetic-order calibration}

Exchange couplings extracted from the predicted Hamiltonian with TB2J\cite{he2021tb2j,liechtenstein1987} agree with couplings extracted in the same way from converged DFT+U Hamiltonians. Over 76 bonds in five calibration materials the correlation is 0.92, the mean absolute difference 1.38 meV and the sign agreement 89.5\% (Fig. 4a). Both sets of couplings are extracted from the same ferromagnetic reference configuration; Supplementary Fig. S8 gives the comparison per material and for other references. For \ce{LaMnO3} the 24 nearest-neighbour couplings agree with a correlation of 0.9997, with every sign correct.

To test the magnetic order that follows from these couplings, we calibrated it against self-consistent DFT+U on 43 GNoME candidates taken from the top of the model's rankings (Supplementary Table S3). Comparing the DFT+U energies of the ferromagnetic state and several antiferromagnetic configurations identifies the lowest of these configurations for 40 of them. The couplings give a binary ferromagnetic-versus-antiferromagnetic call for twenty of the forty, and 17 of these calls are correct: 85.0\% (Wilson\cite{wilson1927} 95\% interval 64.0 to 94.8\%) at an assertion coverage of 50.0\%. The other twenty comprise nineteen abstentions by design and one non-magnetic candidate called ferromagnetic (Supplementary Note 7).

\subsection*{DFT+U validation of altermagnet candidates}

We ran the model on GNoME structures\cite{merchant2023} in their predicted magnetic configurations and adjudicated with DFT+U the 24 altermagnet candidates with the largest predicted splitting, under rules fixed before any DFT+U result was seen (Fig. 5). A candidate is validated when its ferromagnetic state and AFM1, the first antiferromagnetic configuration generated deterministically from its structure, both converge, AFM1 lies below the ferromagnetic state, the spin group of the transition-metal sublattice is d-wave altermagnetic and DFT+U confirms the splitting. This primitive-cell test compares AFM1 only with the ferromagnetic state, so for every candidate that passed it we also converged two antiferromagnetic configurations of a doubled supercell, each against AFM1 in the same supercell (Methods). Five candidates pass every test: \ce{LuCo2B2Ir}, \ce{SrTi2Sb2O}, \ce{Hf3Cr2TcP3}, \ce{HfZr(VRe)2} and \ce{Zr4Mn6AlOs}. In these five, AFM1 lies 77 to 1,116 meV per cell below the ferromagnetic state, stays lowest in the supercell by 83 to 1,169 meV per cell and splits by 393 to 972 meV over the Brillouin zone (469.2 meV for \ce{Hf3Cr2TcP3}, Fig. 5a; Supplementary Table S4). The ground state of these five is thus assigned among the configurations compared: the ferromagnetic state, AFM1 and the two antiferromagnetic configurations of the doubled supercell. In four more candidates DFT+U confirms the predicted splitting, 607 to 1,228 meV, in the antiferromagnetic state. In three of them, \ce{NdLuAl2(Co3B)2}, \ce{NdCd(NiPt)2} and \ce{NdIn(NiPt)2}, the Nd 4f moment is uncompensated, so the full state is ferrimagnetic and the splitting is altermagnetic only on the transition-metal sublattice; the ferromagnetic and supercell references of all four are given in Supplementary Note 7. The outcomes of the other 15 candidates are listed in Supplementary Table S4 and discussed in Supplementary Note 7.

For \ce{Ag3RuO4} we repeated the check at every U. Its DFT+U antiferromagnetic state lies 716 to 793 meV per cell below the ferromagnetically seeded state from U = 0 to 10 eV, with a ruthenium moment of 1.41 to 1.48 $\mu_\mathrm{B}$. Spin-group analysis\cite{smolyanyuk2024amcheck} permits altermagnetism in space group $P4_122$, and at U = 4 eV the DFT+U bands split by 69 meV within 1 eV of the chemical potential, against 67 meV in the model (Fig. 5b). In DFT+U, \ce{Ag3RuO4} is therefore altermagnetic in this antiferromagnetic state, which stays below the ferromagnetically seeded state from U = 0 to 10 eV. The conventional antiferromagnets \ce{NiO} and \ce{MnO} keep their ground states in DFT+U, and their model splittings of 8 and 30 meV lie below the band errors of 139 and 70 meV, so the screening gate, which accepts a predicted splitting only above three times the band error, rejects both (Supplementary Fig. S2a).

\section*{Discussion}

U-HamGNN learns how the Hamiltonian of a magnetic crystal depends jointly on its collinear configuration and on U, for crystals across the periodic table. The spins do not enter the equivariant representation network directly: it predicts the magnetic interaction parameters, couplings J and weights W, and a physics-based assembly builds the two spin channels from them and the input spins (Fig. 1b). Across 71 elements our model reaches a mean absolute error of 1.8 meV on the nonzero matrix elements of the test graphs. Followed across U, it reproduces DFT+U: the gap of \ce{MnTe} opens with U, its spin channels split away from the high-symmetry lines by the amounts DFT+U gives, and the splitting of \ce{Ag3RuO4} follows DFT+U across six U values, including one beyond the trained window. Couplings read from the predicted Hamiltonian agree bond by bond with couplings from converged DFT+U, and the magnetic order they imply is right in 17 of 20 binary calls on a calibration set of GNoME candidates.

A DFT+U survey repeats a self-consistent calculation for every configuration and every U. With a loaded model the same grid costs one query per point: for the 32-atom \ce{Ag3RuO4} cell, 30.4 s on one H100 GPU, 2.5 s of it in the forward pass, against 316 s for one self-consistent DFT+U calculation on 55 CPU cores. The saving comes from reusing one loaded model across many queries (Supplementary Fig. S9), and DFT+U can be spent where it decides the outcome. Among the 24 GNoME altermagnet candidates the model ranked highest, DFT+U confirmed five altermagnets with splittings of 393 to 972 meV, each lowest in both the primitive cell and a doubled supercell, and the predicted splitting of four more. The model can therefore prioritize altermagnet candidates for self-consistent DFT+U adjudication.

Our model extends the universal Hamiltonian models\cite{zhong2024universal,wang2024universal,zhong2026unihamgnn}, which established training across the periodic table and deployment on databases, to the magnetic state and U. Single-family magnetic models\cite{li2023xdeeph,zhong2023tenne3,zhong2023hamgnnpp,yu2024spingnn} learn the dependence on the spin configuration but keep U fixed (Supplementary Note 5).

Our model returns the Hamiltonian of any collinear configuration it is given; whether that configuration survives self-consistency, and which one is the ground state, is settled by DFT+U or an energy comparison, which is why every candidate goes through a DFT+U check. Supplementary Note 7 gives the full account of the remaining candidates and the boundaries of the comparison.

Combining the magnetic term with the spin-orbit coupling of Uni-HamGNN would bring relativistic splitting and magnetic anisotropy into the same framework.

\section*{Methods}

\subsection*{Model architecture}

U-HamGNN predicts the collinear, spin-polarized DFT+U Hamiltonian in the OpenMX basis of numerical atomic orbitals. For each spin channel $\sigma$ the Hamiltonian consists of blocks $H^{\sigma}_{ij}(\mathbf{R})$ between the orbitals of atom $i$ in the home cell and of atom $j$ in the cell at lattice vector $\mathbf{R}$. The bands follow from the generalized eigenproblem $H^{\sigma}(\mathbf{k})\,\mathbf{c} = \varepsilon\, S(\mathbf{k})\,\mathbf{c}$, where $H^{\sigma}(\mathbf{k}) = \sum_{\mathbf{R}} e^{i\mathbf{k}\cdot\mathbf{R}} H^{\sigma}(\mathbf{R})$ and the overlap $S(\mathbf{k})$ is formed in the same way. A rotation $Q$ of the crystal transforms a block as $D^{l_i}(Q)\, H_{ij}\, D^{l_j}(Q)^{\dagger}$ and inversion multiplies it by $(-1)^{l_i+l_j}$; the HamGNN construction\cite{zhong2023hamgnn} satisfies both by assembling every block from irreducible representations (below).

The Hamiltonian of a magnetic crystal also depends on the local moments, and for noncollinear moments it must in addition be equivariant under spin rotation and time reversal. Fitting all spin blocks directly would take eight sub-blocks with different time-reversal parity. HamGNN++\cite{zhong2023hamgnnpp} instead derives the spin-dependent part from a generalized Heisenberg model. The moment of atom $i$ is defined through a weight operator $\hat W_i$, $M_i^{\alpha} = \mathrm{Tr}(\hat W_i\, \hat\sigma^{\alpha}\, \hat\rho)$, with $\hat\rho$ the density matrix and $\hat\sigma^{\alpha}$ the Pauli matrices, and the spin energy is

\begin{equation}
E_{\mathrm{spin}} = \sum_{i<j} \sum_{\alpha\beta} M_i^{\alpha} J_{ij}^{\alpha\beta} M_j^{\beta} + \sum_{k} \sum_{\alpha\beta} M_k^{\alpha} A_k^{\alpha\beta} M_k^{\beta},
\end{equation}

where $J_{ij}^{\alpha\beta}$ and $A_k^{\alpha\beta}$ are the exchange and single-ion tensors. The magnetic part of the Hamiltonian is the variational derivative of $E_{\mathrm{spin}}$ with respect to $\hat\rho$. Each of its terms is linear in one moment and proportional to a matrix element of $\hat W$ and a Pauli matrix. The tensors $J$ and $A$ are even under time reversal and inversion, so the magnetic part changes sign with the moments. The construction is therefore equivariant by design and needs about an eighth of the parameters of a direct fit of all spin blocks.

For collinear moments along $z$ only $\hat\sigma^{z}$ survives, and it is diagonal in spin. The derivative then adds a term $+\Delta$ to the spin-up channel and $-\Delta$ to the spin-down channel, $\Delta_{X} = W_{X} \circ \sum_{a} J_{X,a}\, s_a$ as in the Results. Here $s_a \in \{+1, -1, 0\}$ is the sign of the moment of magnetic atom $a$ (zero for atoms outside the 22-element magnetic allowlist), $J_{X,a}$ are the on-site and pair couplings that reach block $X$ (the $zz$ components of $J$ and $A$, one scalar per pair of orbital shells) and $W_X$ holds the learned elements of the weight operator. Because $\Delta$ is linear in $s$, and $H^{0}$ and $H^{\mathrm{nm}}$ do not depend on the spins, reversing every spin exchanges the two channels exactly, $H^{\uparrow}[-\mathbf{s}] = H^{\downarrow}[\mathbf{s}]$, which is the collinear form of time-reversal equivariance. HamGNN++ also gives the magnitudes of the moments to the network. U-HamGNN gives it only their signs, because the magnitudes are an outcome of the self-consistent calculation rather than a setting of the query, so their effect is left to the learned couplings.

DFT+U adds to the Kohn–Sham potential of the correlated shell of each atom the Dudarev term

\begin{equation}
V^{\sigma}_{mm'} = U_{\mathrm{eff}} \Big( \tfrac{1}{2}\,\delta_{mm'} - n^{\sigma}_{mm'} \Big),
\end{equation}

where $n^{\sigma}$ is the occupation matrix of that shell. It pushes occupied and empty orbitals apart by about $U_{\mathrm{eff}}$, and through the self-consistent occupations it changes both the spin-independent part of the Hamiltonian and the exchange splitting. The reference $H^{0}$ contains no Hubbard term, because OpenMX adds it only from the second self-consistent step, and its two spin channels are identical, so the whole dependence on U and on the spins has to be learned. The U of each atom therefore enters its node attribute (Results), from where it reaches $H^{\mathrm{nm}}$, $J$ and $W$ through every layer. HamGNN++ was trained on individual magnetic systems (Fe/Ir(111), NiBr$_2$ and NiBrI) with noncollinear moments and spin-orbit coupling. U-HamGNN keeps its parameterization in the collinear form, has no spin-orbit term, is trained across the periodic table and takes U as an input.

U-HamGNN uses the Uni-HamGNN equivariant architecture\cite{zhong2026unihamgnn} and is trained on the magnetic DFT+U data described below. The model learns the self-consistent correction to a reference Hamiltonian (delta learning). It keeps the HamGNN construction\cite{zhong2023hamgnn} on the irreducible representations (irreps) of the e3nn framework\cite{geiger2022e3nn,thomas2018tfn,batzner2022nequip}, with the OpenEquivariance strict-mode kernel backend\cite{bharadwaj2025oeq}. The representation network takes a crystal graph with a 26.0 Bohr (13.76 Å) cutoff and a cosine cutoff function. Every atom carries a 116-dimensional scalar attribute. It concatenates a one-hot encoding over 84 atomic types (H to Po, more than the 71 elements present in the training data) with a 32-function Gaussian expansion of the effective on-site interaction $U_i$ on [0, 8] eV (width $\sigma_U = 0.26$ eV), from which the expansion of $U = 0$ is subtracted; an atom without a Hubbard correction therefore reduces to the plain element embedding (zero-centred U embedding). Edges carry spherical harmonics up to $l = 6$ (component normalization) and 128 Bessel radial functions. A pair-interaction embedding builds the initial edge features from the two atom attributes, the spherical harmonics and a radial multilayer perceptron with layers [256, 256, 128]. An atom-wise linear layer lifts the attributes to equivariant node features with irreps 64x0e+64x0o+48x1o+48x1e+48x2o+48x2e+32x3o+32x3e+16x4o+16x4e+8x5o+8x5e+4x6e (1,860 components). Three interaction layers follow, each made of an equivariant convolution and a pair-interaction block. The convolution forms a fully connected tensor product of the sender features with the edge spherical harmonics, scales it by radial weights from the radial network, sums over neighbours and passes the result through a gated nonlinearity with a residual connection and a linear skip path. The pair-interaction block then updates the edge features from the two updated node features. The magnetic state enters only in the output network.

The output network (OpenMX basis, at most 26 orbitals per atom, s3p2d2f1) reads orbital blocks from the node and edge features. Each $l_a l_b$ shell pair is decomposed into irreps $L = |l_a - l_b|, \ldots, l_a + l_b$ (134 irreps, 676 components) and reassembled with Clebsch–Gordan coefficients, $H^{(l_a l_b)}_{m_a m_b} = \sum_{L} \sqrt{2L+1} \sum_{M} \langle l_a m_a\, l_b m_b | L M\rangle\, c^{(l_a l_b L)}_{M}$, then reordered to the OpenMX orbital order and masked to the orbitals present on each element. On-site blocks are projected onto the Hermitian subspace, and off-site blocks are made Hermitian against the reverse edge, $H_{ij}(\mathbf{R}) = H_{ji}(-\mathbf{R})^{\dagger}$. Six heads share this decoder: on-site and off-site heads for the spin-independent Hamiltonian ($H^{\mathrm{nm}}$), for the exchange (one scalar coupling $J$ per shell pair) and for the weights (a full 26 × 26 matrix $W$). The collinear magnetic term uses the input spin directions $s_i \in \{+1, -1, 0\}$ of the atoms in the 22-element magnetic allowlist (Ti, V, Cr, Mn, Fe, Co, Ni, Cu, Mo, Tc, Ru, Rh, Pd, Ce, Nd, Sm, Dy, Ho, Re, Os, Ir, Pt). The on-site term of magnetic atom $i$ is $W_i \circ (J_i s_i + \sum_{j \to i} J_{ij} s_j)$, where the sum runs over the neighbouring magnetic atoms through the off-site couplings; the off-site term of bond $ij$ at lattice vector $\mathbf{R}$ is $W_{ij} \circ (J_i s_i + \delta_{\mathbf{R},0} J_j s_j + \sum_{k} J_{ik} s_k)$ with the same neighbour sums. The two spin channels are $H^{\uparrow} = H^{0} + H^{\mathrm{nm}} + \Delta$ and $H^{\downarrow} = H^{0} + H^{\mathrm{nm}} - \Delta$. $H^{0}$ is the non-self-consistent reference Hamiltonian that the OpenMX post-processing tool writes from the superposition of atomic densities. It does not depend on spin or U, comes with every graph and lets the network learn the self-consistent correction. During training, band energies are computed by generalized diagonalization $L^{-1} H(\mathbf{k}) L^{-\dagger}$, with $L$ the Cholesky factor of the overlap $S(\mathbf{k})$, at four automatically chosen k points for the 8 bands above and below the valence-band maximum in each spin channel, after an overlap-weighted alignment of the energy zero between prediction and reference. The gradients of the exchange and weight heads are scaled by 15 relative to the Hamiltonian head. The model has 23,650,400 trainable parameters (23.26 million in the representation network, 0.39 million in the output heads). It is collinear and has no spin-orbit term.

\subsection*{Training data and split}

The Hamiltonians were computed with OpenMX\cite{ozaki2003,ozaki2004} DFT+U (Dudarev form, $U_{\mathrm{eff}} = U - J$, written to the OpenMX Hubbard block\cite{dudarev1998,han2006}) for magnetic crystal structures from the Materials Project\cite{jain2013} and MAGNDATA-derived magnetic cells\cite{gallego2016}. The full LMDB holds 205,251 graphs over 42,744 distinct structures. A structure is stored at up to three U settings (0 eV, the tabulated value and an 8 eV ceiling) and in two to six collinear magnetic states; 90.8\% of the structures appear at more than one U and 91.9\% in more than one magnetic state. Every graph carries three per-atom fields: \texttt{spin\_length} (magnitude of the Mulliken moment, $\mu_\mathrm{B}$), \texttt{spin\_vec} (its unit direction, of which the collinear model uses the sign of the z component) and \texttt{u\_atoms} ($U_{\mathrm{eff}}$ in eV, zero where no correction is applied). We removed the 47,551 graphs in which any allowlisted magnetic atom has a moment below 0.5 $\mu_\mathrm{B}$. The remaining 157,700 graphs were shuffled with a fixed seed and split 98/1/1 into 154,546 training, 1,577 validation and 1,577 test graphs. The training subset covers 35,632 structures and 71 elements; Supplementary Table S2 gives per-element structure and graph counts and Supplementary Table S6 the full configuration. The test split was drawn at the graph level, so it contains configurations of structures whose other U or spin configurations appear in training. This is the relevant test for an operator conditioned on U and spin, which has to predict the Hamiltonian at a given U and magnetic state.

\subsection*{Training protocol}

Training ran on 128 NVIDIA H100 GPUs (16 nodes × 8) with one graph per GPU. Stage 1 fitted the Hamiltonian only, with a Huber loss ($\delta = 5 \times 10^{-4}$ Ha) on the Hamiltonian blocks and a spin-splitting auxiliary loss on $H^{\uparrow} - H^{\downarrow}$ (weights 10,000 and 150; the auxiliary loss is restricted to blocks with a nonzero reference splitting). It used the spectral-hybrid optimizer at a learning rate of 0.02, reduced to 0.005 on resumption from epoch 26, and ran to epoch 138 with a plateau scheduler (factor 0.5). Stage 2 continued from the last Stage 1 checkpoint with an L1 Hamiltonian loss (weight 27,000.211) and an L1 band-energy loss (weight 272.11, ramped linearly from zero over 20 epochs). It used a learning rate of 0.001 for both optimizer groups and gradient clipping at 100, and the validation band-energy error drove the scheduler, early stopping and checkpoint selection. The best checkpoint is Stage-2 epoch 78 (global step 95,432, counted from the start of Stage 2; validation Hamiltonian MAE 2.092 meV, validation band MAE 96.5 meV). It is used for every result in this paper. Fig. 2b shows the per-epoch validation curves of both stages.

\subsection*{Spectral-hybrid optimizer}

For a matrix parameter $W \in \mathbb{R}^{n \times m}$ with gradient $G_t$ the optimizer keeps a momentum, an exponential average of past gradients and their difference, a temporal innovation that gives a cheap replica of the gradient noise\cite{yaida2019fdt},

\begin{equation}
M_t = \beta M_{t-1} + (1-\beta) G_t, \qquad \bar G_t = \beta_m \bar G_{t-1} + (1-\beta_m) G_t, \qquad \xi_t = G_t - \bar G_{t-1}.
\end{equation}

The momentum starts from $M_0 = 0$ and the average from $\bar G_0 = G_1$, so the two differ although $\beta = \beta_m$; the innovation $\xi_t$ is used only by the metric variant described below. The optimizer

refreshes a rank-$r$ basis by one power iteration on the momentum, a persistent basis that replaces the Newton–Schulz orthogonalization of Muon\cite{liu2025muon,bernstein2024norm},

\begin{equation}
V_t = \mathrm{qf}\!\big(M_t^{\mathsf T} M_t V_{t-1}\big), \qquad Y_t = M_t V_t = U_t\,\mathrm{diag}(\sigma_t),
\end{equation}

where qf is the Q factor of a reduced QR decomposition, $\sigma_t$ holds the column norms of $Y_t$ and $U_t$ is $Y_t$ with its columns scaled to unit norm (it is not orthogonalized again),

and, every $p$ steps (the spectral period), projects the gradient and its replica onto the basis and compares them direction by direction,

\begin{equation}
\begin{aligned}
g^{(1)} &= \mathrm{diag}\!\big(U_t^{\mathsf T} G_t V_t\big), \quad g^{(2)} = \mathrm{diag}\!\big(U_t^{\mathsf T} \bar G_{t-1} V_t\big),\\
B &= 1 + \lambda_\nu\,\nu + \lambda_c\Big(1 - \tfrac{2 g^{(1)} g^{(2)}}{g^{(1)2} + g^{(2)2} + \epsilon}\Big), \quad \nu = \tfrac{1}{4}\big(g^{(1)} - g^{(2)}\big)^2,
\end{aligned}
\end{equation}

so that $B$ grows where the two estimates disagree. The spectral weights are a softmax over the per-direction advantage $a = \eta\,\tfrac{1}{2}(g^{(1)} + g^{(2)}) - \eta^2 B$, projected onto a tube around unit weight,

\begin{equation}
w = \Pi_{\delta}\!\Big[\sqrt{r\,\mathrm{softmax}\big(\beta_w\, z(a)\big)}\Big], \qquad \|w\|_2 = \sqrt{r},
\end{equation}

where $\beta_w$ is chosen on a grid by maximizing the cross-fitted surrogate gain defined below. If no temperature gives a positive gain, $w = 1$ and the step reduces to streaming Muon. The update is

\begin{equation}
W_{t+1} = W_t - \eta\, c\, U_t\, \mathrm{diag}(w)\, V_t^{\mathsf T}, \qquad c = \sqrt{\min(n,m)/nm}\,\big/\,\mathrm{rms}\big(U_t\,\mathrm{diag}(w)\,V_t^{\mathsf T}\big),
\end{equation}

with the root-mean-square normalization of Muon. These equations apply to every parameter on the spectral route, and the rest take an AdamW step\cite{loshchilov2019adamw,kingma2015adam}. Parameters are routed by role and shape. A tensor goes to the spectral route if it has at least two dimensions and its parameter name contains none of the strings embedding, radial, basis, \texttt{output\_module}, norm or bias. The production model stores the e3nn tensor-product and linear weights as flattened one-dimensional tensors, and the parameter names of the multilayer perceptrons that map the radial basis to the tensor-product weights contain none of these strings, so the spectral route holds the 48 dense matrices of these perceptrons (14,186,496 parameters, 60\%) and the AdamW group holds the remaining 80 tensors (9,463,904 parameters). Constants: momentum $\beta = 0.95$, gradient-EMA $\beta_m = 0.95$, rank $r = 32$, one power iteration, spectral refresh every $p = 4$ steps, $\lambda_\nu = 2.0$, $\lambda_c = 0.25$, tube half-width $\delta = 0.35$, temperature grid $\beta_w \in \{0, 0.001, 0.002, 0.005, 0.01, 0.02, 0.05, 0.1, 0.25, 0.5, 1, 2, 4\}$, surrogate penalties $\lambda_{\mathrm{diff}} = 0.5$, $\lambda_{\mathrm{unc}} = 0.5$, $\lambda_{\mathrm{tube}} = 0.05$, hard selection below a learning rate of 0.05, no weight decay; AdamW uses $\beta_1 = 0.99$, $\beta_2 = 0.999$, $\epsilon = 10^{-8}$. The tube projection clamps the weights to $[1 - \delta, 1 + \delta]$ and renormalizes $\|w\|_2 = \sqrt{r}$ (all constants and surrogate definitions are collected in Supplementary Note 6); $z(a)$ standardizes the advantage vector to zero mean and unit root-mean-square. The cross-fitted surrogate gain evaluates $\eta\, g^{(2)} \cdot w^{(1)} - \tfrac{1}{2}\eta^2 B \cdot w^{(1)2}$ for weights $w^{(1)}$ fitted on $g^{(1)}$, and symmetrically, relative to unit weights; its penalties are the $\nu$-weighted mean of $(w^2 - 1)^2$ (diffusion), the absolute difference of the two cross gains (uncertainty) and the mean squared distance from unit weight (tube). The metric variant keeps exponential averages of the row and column energies of the gradient EMA (signal, $S$) and of the innovation (noise, $N$), forms the reliability $\rho = S/(S + N)$, sets row and column temperatures $T \propto \rho^{1/4}$ normalized to unit geometric mean and clipped to [0.5, 2], builds the basis and the projections on $\sqrt{T_r}\, M\, \sqrt{T_c}$ and removes the metric before the update. The Muon baseline uses momentum 0.95 with Nesterov mixing, five quintic Newton–Schulz iterations with coefficients (3.4445, $-$4.7750, 2.0315) and the scale $\max(1, \sqrt{n/m})$.

\subsection*{Optimizer benchmarks}

Both optimizer benchmarks used the same 500 iron-containing graphs from the training LMDB (Fe-500), with a fixed 500/100/100 split, 10 epochs of 500 steps at batch size one, strict OpenEquivariance kernels and the global random seed 666, and report the Hamiltonian MAE on the 100 validation graphs at the end of training. The comparison of Fig. 2a used a 33.2-million-parameter configuration. AdamW ran at a learning rate of 3.2 × $10^{-3}$. Muon updated the same matrix parameters as the spectral route at a learning rate of 0.02, with AdamW at 3.2 × $10^{-3}$ for the other tensors, and was launched three times with the same seed (25.679, 25.682 and 25.676 meV; Fig. 2a shows the first launch). The spectral-hybrid optimizer used the same parameter split with period 4, rank 32, a spectral learning rate of 0.02 and an AdamW learning rate of 5.6 × $10^{-3}$. The matched benchmark used a 14.2-million-parameter configuration (node irreps starting 32x0e+32x0o, 64 Bessel functions) in which every optimizer ran at one learning rate of 3.2 × $10^{-3}$ and, for Muon and the spectral update, on every tensor with at least two dimensions; the spectral update used a full-rank basis refreshed every step. AdamW, Muon, the spectral update and its metric variant were each run with seeds 666, 1 and 2 in otherwise identical configurations. The two benchmarks and the production run use three different model sizes and are reported separately (Supplementary Table S1).

\subsection*{Test evaluation and band-structure comparison}

Test errors are computed over the matrix elements whose DFT+U reference is nonzero. Each orbital block is stored as a 2 × 26 × 26 array, and entries outside the basis of its two atoms are zero padding in both the reference and the prediction; they are excluded, which leaves 1,482,157,610 of the 5,506,865,000 stored entries of the 4,073,125 spin-resolved test blocks. The MAE and coefficient of determination in Fig. 2c are pooled over all of these elements (1.802 meV, 0.99984), and the per-graph MAE averages over the nonzero elements of one graph (mean 2.047 meV, median 1.667 meV, 95th percentile 4.302 meV). Per-element errors attribute each graph's MAE to every element in its formula, so they are formula-attributed, not orbital-block-resolved. The band structures in Figs. 3 and 5 and Supplementary Fig. S1 come from diagonalizing the predicted and the converged DFT+U Hamiltonian on the same 141-point k path (protocol and all window errors in Supplementary Note 3 and Supplementary Table S8). In this comparison the structure is fixed at every U (no relaxation), and the model predicts the Hamiltonian for the magnetic state of the converged DFT+U calculation. Both sides are referenced to the OpenMX chemical potential of the converged calculation, and the window MAE is the mean absolute eigenvalue difference over all band and k-point pairs where either side lies within 1 eV of it. The model and DFT+U are measured with the same definitions. The path gap is zero when any band crosses the chemical potential; otherwise it is the difference between the lowest eigenvalue above and the highest below it, over both spin channels. The same-band splitting is the largest $|E_{n\uparrow}(\mathbf{k}) - E_{n\downarrow}(\mathbf{k})|$ over the plotted path, pairing bands by index within each spin channel and keeping pairs with both energies within 2 eV of the chemical potential (1 eV where stated; Supplementary Table S8 gives the 0.5, 1, 2 and 4 eV values). The candidate ranking used a different, model-only quantity, the largest splitting on an 11 × 11 × 10 Brillouin-zone mesh within 4 eV of an estimated Fermi level with bands paired by nearest energy (168.1 meV for \ce{Ag3RuO4} at U = 4 eV). It ranks candidates and is not used as a validation number. The adjudication splitting (Fig. 5a,c and Supplementary Table S4) is a third quantity, measured over a 9 × 9 × 9 Brillouin-zone mesh within 4 eV of the estimated Fermi level: the largest distance from an eigenvalue of one spin to the nearest eigenvalue of the other spin, which needs no band pairing. It was computed with the same code and settings from the converged DFT+U Hamiltonian of the antiferromagnetic state and, for the model values in Fig. 5c, from the predicted Hamiltonian of that state. The test band overlays (Supplementary Fig. S5) solve the generalized eigenproblem on the stored predicted and reference Hamiltonians with the stored overlap matrices, at four path nodes and twelve off-axis points per graph.

For the counterfactual test (Supplementary Fig. S4d) we drew, with a fixed seed (20261002), 200 of the 204 test graphs that are antiferromagnetic-like and contain a magnetic atom with 0 < U < 8 eV. Three copies changed only the conditioning inputs (every U set to 0; every U between 0 and 8 eV set to 8 eV; every spin set along +z) and were evaluated with the same checkpoint, runtime settings and metric code. The stored DFT+U targets and $H^{0}$ were left unchanged while only the conditioning inputs were altered. The unmodified copy reproduces the original test errors exactly.

\subsection*{DFT+U reference calculations}

All first-principles references were computed with OpenMX 3.9 full self-consistent SCF with the Dudarev $U_{\mathrm{eff}}$ correction\cite{dudarev1998,han2006}. The application uses a 45-element $U_{\mathrm{eff}}$ table (31 d-channel and 14 f-channel entries, Supplementary Table S5 and Supplementary Fig. S7). Its 8 entries for La, Lu, Tb, Dy, Ho, Er, Tm and Yb exceed the trained window of 8 eV and are extrapolations; the 43-candidate calibration uses only the in-window d-channel subset ($U_{\mathrm{eff}} \le 6.4$ eV). For each calibration candidate we converged a set of magnetic configurations, the ferromagnetic configuration plus antiferromagnetic configurations generated deterministically from the structure, on fully scaled cells; the first antiferromagnetic configuration in this order is AFM1. The ground state is the lowest configuration after normalization to the parent cell, and a near-degeneracy flag comes from the per-magnetic-atom gap to the opposite-label competitor (gate at 1.0 meV). Supplementary Table S3 is the full ledger and Supplementary Table S4 lists the adjudicated candidates. The 43 candidates are GNoME structures from the top of the model's rankings, and their reduced compositions do not occur among the 24,774 reduced compositions of the training structures (pymatgen\cite{ong2013pymatgen} check). \ce{Ag3RuO4} and the controls are Materials Project compounds whose compositions do occur in training, and they are reported separately for that reason.

\subsection*{Validation criteria and controls}

A screening gate fixed before any DFT+U result was seen accepts a screened altermagnet candidate only with a self-consistent antiferromagnetic ground state and sustained moment, symmetry-permitted altermagnetism and a same-band splitting at least three times the model's band error on that structure; it decides which candidates are prioritized. The validation verdict uses DFT+U quantities only: both magnetic states converged with the antiferromagnetic state lowest, a sustained moment, a spin-group classification that permits altermagnetism\cite{smolyanyuk2024amcheck}, and a nonzero spin splitting of the DFT+U bands of the antiferromagnetic state, measured as defined above. A candidate whose antiferromagnetic splitting is confirmed but whose ground state could not be settled is recorded as partial. The 24 GNoME candidates were run with the tabulated $U_{\mathrm{eff}}$ of each element, in DFT+U and in the model alike.

The ferromagnetic references were first started from the occupations of the screening inputs, which carry no moment. For the 14 candidates whose first reference failed or collapsed they were rerun from polarized moments (the antiferromagnetic magnitudes, all aligned, total occupation per site unchanged) with otherwise identical settings, and with Kerker-preconditioned mixing as a second variant. By a rule fixed before the result, the same rerun was applied to \ce{NdLuAl2(Co3B)2}, whose first ferromagnetic reference had converged only to a low-moment solution. Where the two variants converge to different ferromagnetic solutions, the lower one is the reference. A run counts as converged when the band energy changes by less than $10^{-7}$ Hartree between steps, has moved by less than 1 meV over the five preceding steps and the density residual has fallen at least a thousandfold during the run. A candidate whose rerun converges below the antiferromagnetic state is negative; one whose rerun converges above it, with its moments kept, passes this step; one for which neither variant converges is partial. The convergence record of each rerun is in Supplementary Note 7.

The primitive-cell comparison sets AFM1 only against the ferromagnetic state. For the seven candidates that passed it we therefore converged, under a protocol fixed before the results, the two antiferromagnetic configurations of a doubled supercell generated deterministically from the structure, each against AFM1 in the same supercell. These runs used the k-point density of the primitive 6 × 6 × 6 mesh, the production input, a cold start, the convergence rule above and the same reruns. A candidate stays validated when AFM1 is lowest and beats every other configuration by more than 1.0 meV per magnetic atom and by more than the energy swing of either run over its last five steps. A lower configuration that is not altermagnetic makes it negative; a margin inside these bounds, or a required configuration that stays unconverged after these reruns, makes it partial. AFM1 stayed lowest in five candidates, with the next configuration 83 to 1,169 meV per parent cell higher.

The spin-group class comes from FindSpinGroup\cite{yu2026findspingroup,liu2022spingroup,jiang2024ssg} on the converged antiferromagnetic state, read on the transition-metal sublattice that carries the splitting, with moment magnitudes averaged over each crystallographic orbit of the non-magnetic structure so that sub-0.1 $\mu_\mathrm{B}$ scatter between equivalent sites does not break the symmetry; where a rare-earth 4f moment is present it is uncompensated and the full state is ferrimagnetic. Two variants of this rule are reported in Supplementary Table S4 and Supplementary Note 7; comparing the multisets of up and down moment magnitudes per element gives the same verdict for all five validated candidates. \ce{NiO} and \ce{MnO} were run through the identical gate as negative controls. \ce{Ag3RuO4} was converged in the ferromagnetically seeded and AFM1 states at $U_{\mathrm{eff}}$ = 0, 2, 4, 6, 8 and 10 eV, compared with the model on the same k path at each U, and classified with spinspg/amcheck (space group $P4_122$, four Ru sites).

\subsection*{Exchange couplings and magnetic order}

Exchange couplings $J_{ij}$ are extracted with TB2J\cite{he2021tb2j} from the predicted Hamiltonian of the ferromagnetic reference configuration (Green's-function formula\cite{liechtenstein1987}, 7 × 7 × 7 k mesh, per-material cutoff). The DFT-side couplings for the parity use the same settings on the converged OpenMX Hamiltonian of the same reference configuration, with bonds paired by sites and lattice vector. Only isotropic couplings are used. The zero-temperature order of each material comes from simulated annealing\cite{kirkpatrick1983} of the isotropic Heisenberg model on the learned couplings (Supplementary Note 2). The candidate rankings ran the frozen model on GNoME structures\cite{merchant2023} in their predicted ground-state configuration, and the spin-group class follows the spin-space-group formalism\cite{liu2022spingroup,jiang2024ssg}.

\subsection*{Computational cost}

Timings refer to the 32-atom \ce{Ag3RuO4} cell at U = 4 eV. The DFT+U reference is the mean of 12 self-consistent OpenMX runs on 55 MPI ranks (316.3 s). The model timing is the warm compute core on one NVIDIA H100 GPU (30.4 s for graph construction, the forward pass and conversion of the predicted blocks to the OpenMX format), of which the forward pass takes 2.5 s. A single wrapped call that includes start-up and model loading took 420.2 s, which is not faster than one DFT+U calculation; the saving requires reusing the loaded model across queries. The comparison spans different hardware and one structure, and it is not a scaling measurement.

\section*{Data availability}

The data that support the findings of this study are available from the corresponding authors on reasonable request.

\section*{Code availability}

The code is available at \url{https://github.com/QuantumLab-ZY/HamGNN}.

\bibliographystyle{naturemag}
\bibliography{references}

\section*{Acknowledgements}

We acknowledge financial support from the National Key Research and Development Program of China (grant no. 2022YFA1402901, H.X.), NSFC (grant no. 12504262, Y.Z.), NSFC (grant no. 12188101, H.X.), Shanghai Science and Technology Program (grant no. 23JC1400900, H.X.), the Guangdong Major Project of the Basic and Applied Basic Research (Future functional materials under extreme conditions—grant no. 2021B0301030005, H.X.), Shanghai Pilot Research—Fudan University 21TQ1400100 (grant no. 23TQ017, H.X.), the robotic AI-Scientist platform of Chinese Academy of Sciences (H.X.) and New Cornerstone Science Foundation (H.X.).

\section*{Competing interests}

The authors declare no competing interests.

\clearpage
\begin{figure}[p]
\centering
\includegraphics[width=\textwidth,height=0.72\textheight,keepaspectratio]{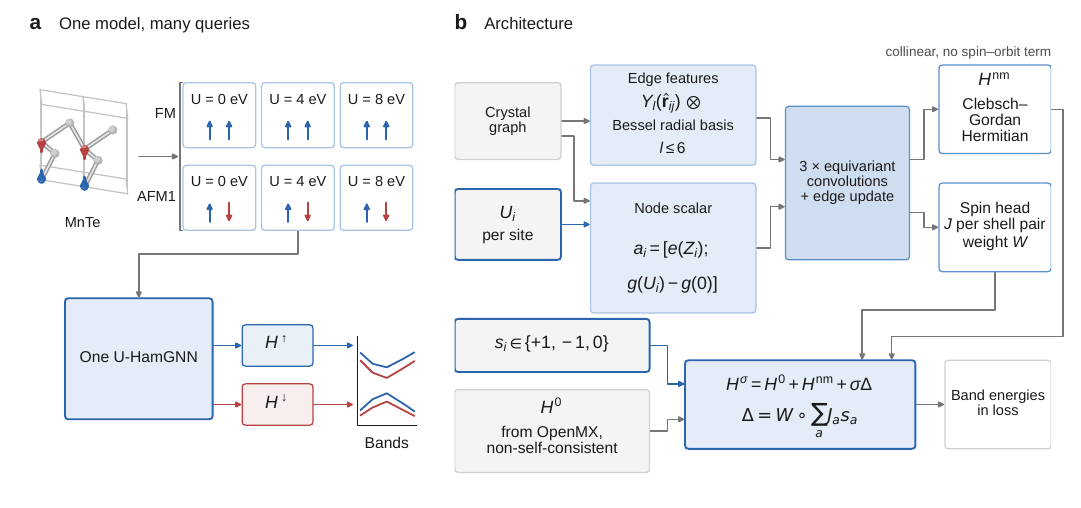}
\caption{\textbf{U-HamGNN: spin and Hubbard U as model inputs.} \textbf{a}, One crystal (\ce{MnTe}) queried with our model in two collinear magnetic configurations (FM and AFM1) at three values of U; each query returns the two spin channels of the Hamiltonian, from which the bands follow by diagonalization. \textbf{b}, Architecture. Each atom carries a scalar attribute $a_i$ that joins its element embedding to a zero-centred Gaussian expansion of its U; edges carry spherical harmonics up to $l = 6$ and a Bessel radial basis. Three equivariant convolution layers with edge updates feed two heads: a spin-independent block $H^{\mathrm{nm}}$ assembled through Clebsch–Gordan coupling and made Hermitian, and a spin head that returns a coupling $J$ per pair of orbital shells and a weight matrix $W$. The input spins $s_i$ and the non-self-consistent OpenMX reference $H^{0}$ enter only at the assembly step, $H^{\sigma} = H^{0} + H^{\mathrm{nm}} + \sigma\Delta$ with $\Delta = W \circ \sum_a J_a s_a$, and band energies enter the training loss. The model is collinear and has no spin-orbit term. The elements of the training set are shown in Supplementary Fig. S6.}
\label{fig:F1}
\end{figure}
\clearpage
\begin{figure}[p]
\centering
\includegraphics[width=\textwidth,height=0.72\textheight,keepaspectratio]{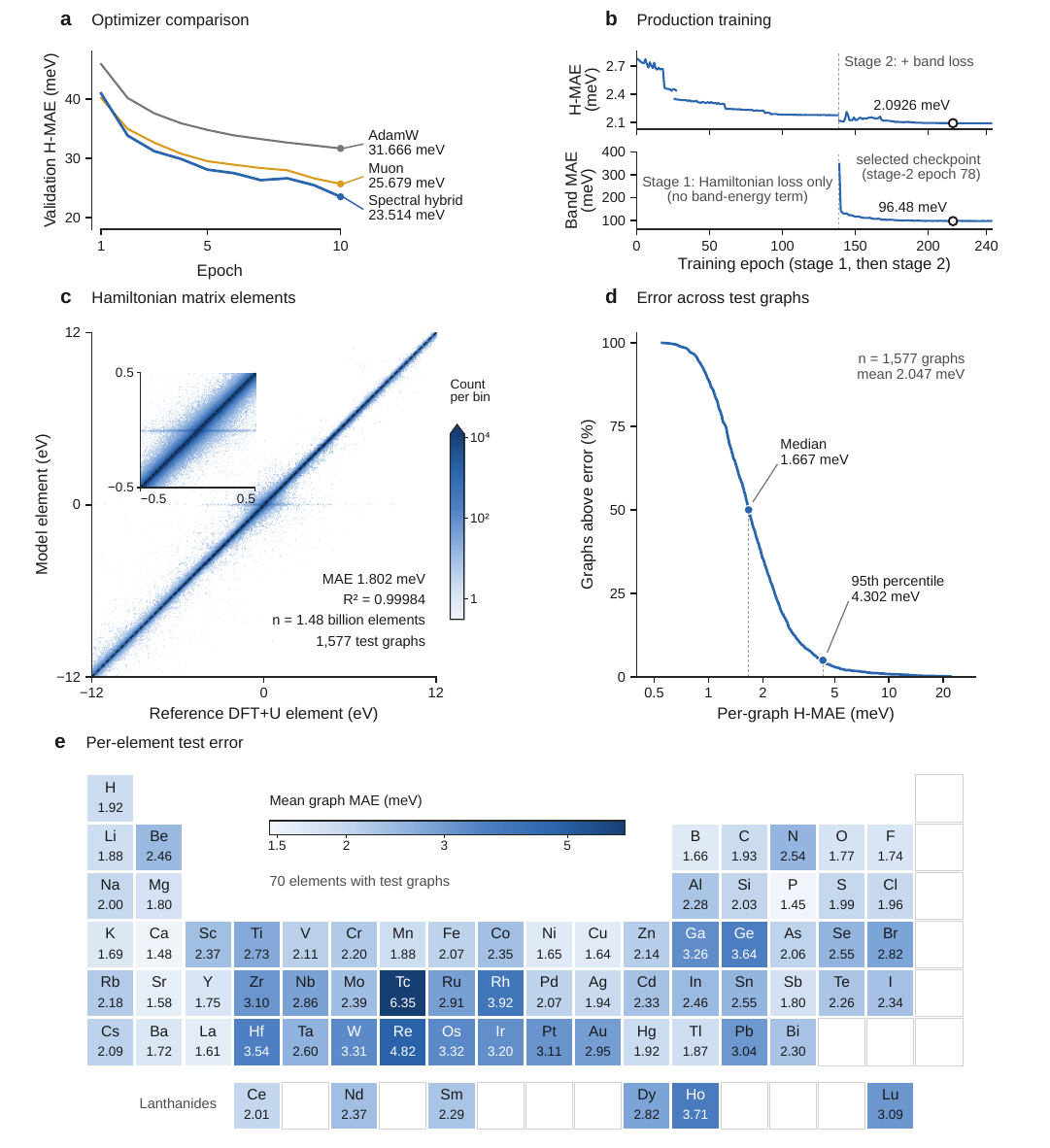}
\caption{\textbf{Spectral-hybrid training and test accuracy.} \textbf{a}, Validation Hamiltonian mean absolute error (MAE) of AdamW, Muon and the spectral-hybrid optimizer over ten epochs on 500 iron-containing graphs (33.2-million-parameter configuration, seed 666); final values at the line ends. Supplementary Fig. S3b gives a matched three-seed comparison. \textbf{b}, Validation Hamiltonian MAE (upper) and validation band MAE (lower) of the production run on one epoch axis, through the Hamiltonian-only stage 1 and the band-energy stage 2 (stage 2 epochs offset to continue the axis; the band loss and its validation error exist only in stage 2); open circles, the selected checkpoint (stage 2 epoch 78), 2.0926 meV and 96.48 meV. \textbf{c}, Predicted against reference DFT+U Hamiltonian elements as a log-scaled two-dimensional count over all 1,482,157,610 nonzero elements of the 1,577 test graphs; dashed line, identity; inset, elements within 0.5 eV of zero. MAE 1.802 meV, $R^2$ 0.99984. \textbf{d}, Fraction of test graphs whose per-graph MAE exceeds a given value; median 1.667 meV, 95th percentile 4.302 meV, mean 2.047 meV. \textbf{e}, Per-element test MAE for the 70 elements with test coverage, attributing each graph's error to every element in its formula (graph-attributed, not orbital-resolved). Errors in c–e are taken over the matrix elements whose DFT+U reference is nonzero, without the zero padding of the orbital blocks (Methods). All values are for the frozen checkpoint.}
\label{fig:F2}
\end{figure}
\clearpage
\begin{figure}[p]
\centering
\includegraphics[width=\textwidth,height=0.72\textheight,keepaspectratio]{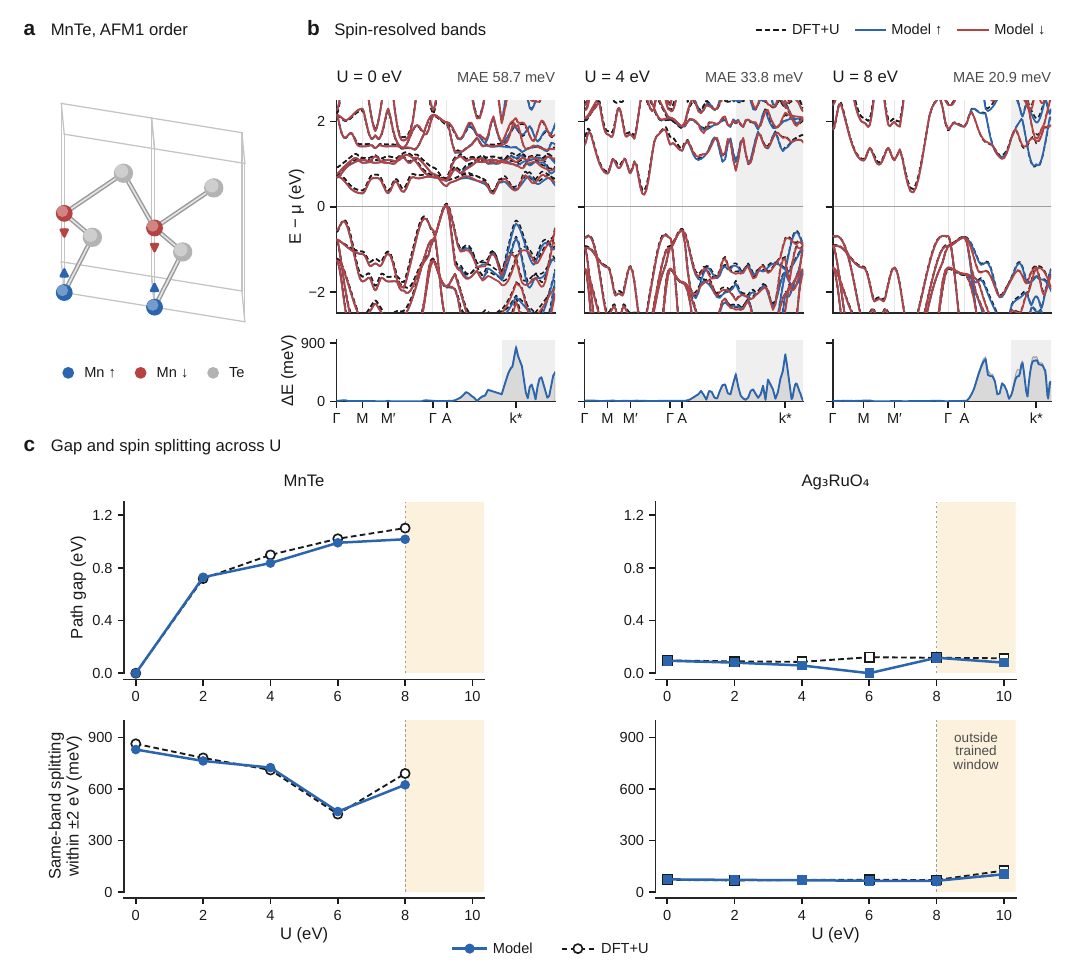}
\caption{\textbf{Band gaps and spin splitting across Hubbard U.} \textbf{a}, \ce{MnTe} in the AFM1 configuration (two repeats of the four-atom cell; arrows, Mn spins). \textbf{b}, Spin-resolved bands at U = 0, 4 and 8 eV at a fixed structure, for the magnetic state of the converged DFT+U calculation, on the same 141-point k path (dashed black, DFT+U; blue and red, model spin up and down), relative to the OpenMX chemical potential of the converged calculation; headers give the band MAE within 1 eV of it (58.7, 33.8 and 20.9 meV). The path runs through the fractional points Γ (0, 0, 0), M (½, 0, 0), M$'$ (½, ½, 0, equivalent to M), Γ and A (0, 0, ½) of the hexagonal cell; shading marks the scan through k* = (0.158, 0, 0.25), the point of largest predicted splitting. Lower strips, largest same-band spin splitting at each k over bands within 2 eV of the chemical potential (fill, DFT+U; line, model). \textbf{c}, Path gap and largest same-band splitting within 2 eV against U for \ce{MnTe} (U = 0 to 8 eV in steps of 2 eV) and \ce{Ag3RuO4} (U = 0 to 10 eV), model (blue, filled) and DFT+U (black, open; circles, \ce{MnTe}; squares, \ce{Ag3RuO4}); each point is one calculation without replicate spread, and the two systems share one y scale per row and one U axis, so model–DFT+U differences can be compared directly. The gap is zero when a band crosses the chemical potential; bands are paired by index within each spin channel. Shading marks U above the trained window of 8 eV.}
\label{fig:F3}
\end{figure}
\clearpage
\begin{figure}[p]
\centering
\includegraphics[width=\textwidth,height=0.72\textheight,keepaspectratio]{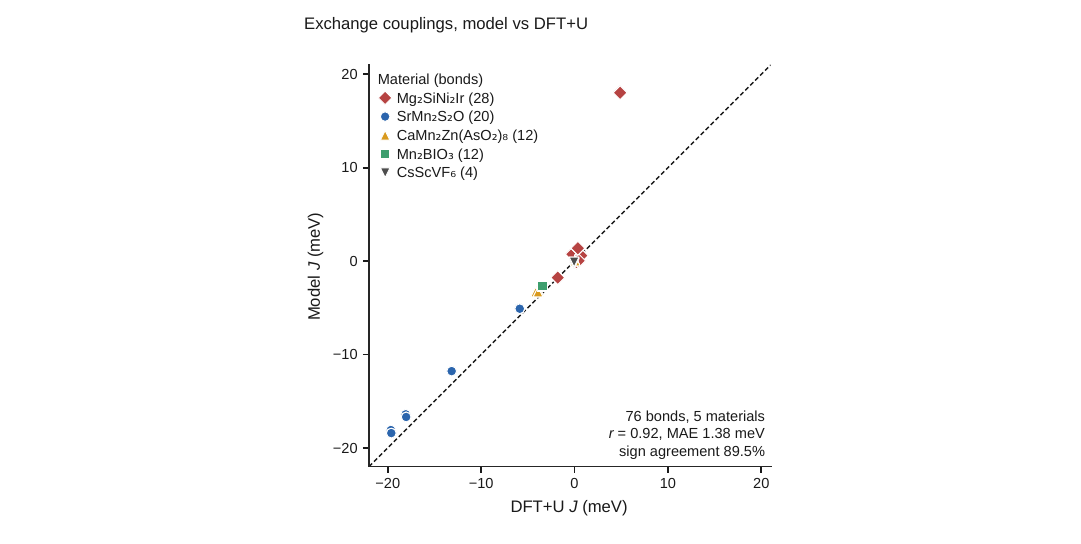}
\caption{\textbf{Exchange couplings, model against DFT+U.} Isotropic exchange couplings extracted with TB2J from the model and from the converged DFT+U Hamiltonian on the same ferromagnetic reference configuration, for 76 bonds in five calibration materials (colour and marker shape identify the material, legend with bond counts; dashed line, identity): Pearson r 0.92, MAE 1.38 meV, sign agreement 89.5\%; the isolated point at the top is four overlapping, symmetry-equivalent shortest bonds of \ce{Mg2SiNi2Ir} (2.49 Å; DFT+U 4.9 meV, model 18.0 meV), which carry about half of the total absolute deviation.}
\label{fig:F4}
\end{figure}
\clearpage
\begin{figure}[p]
\centering
\includegraphics[width=\textwidth,height=0.72\textheight,keepaspectratio]{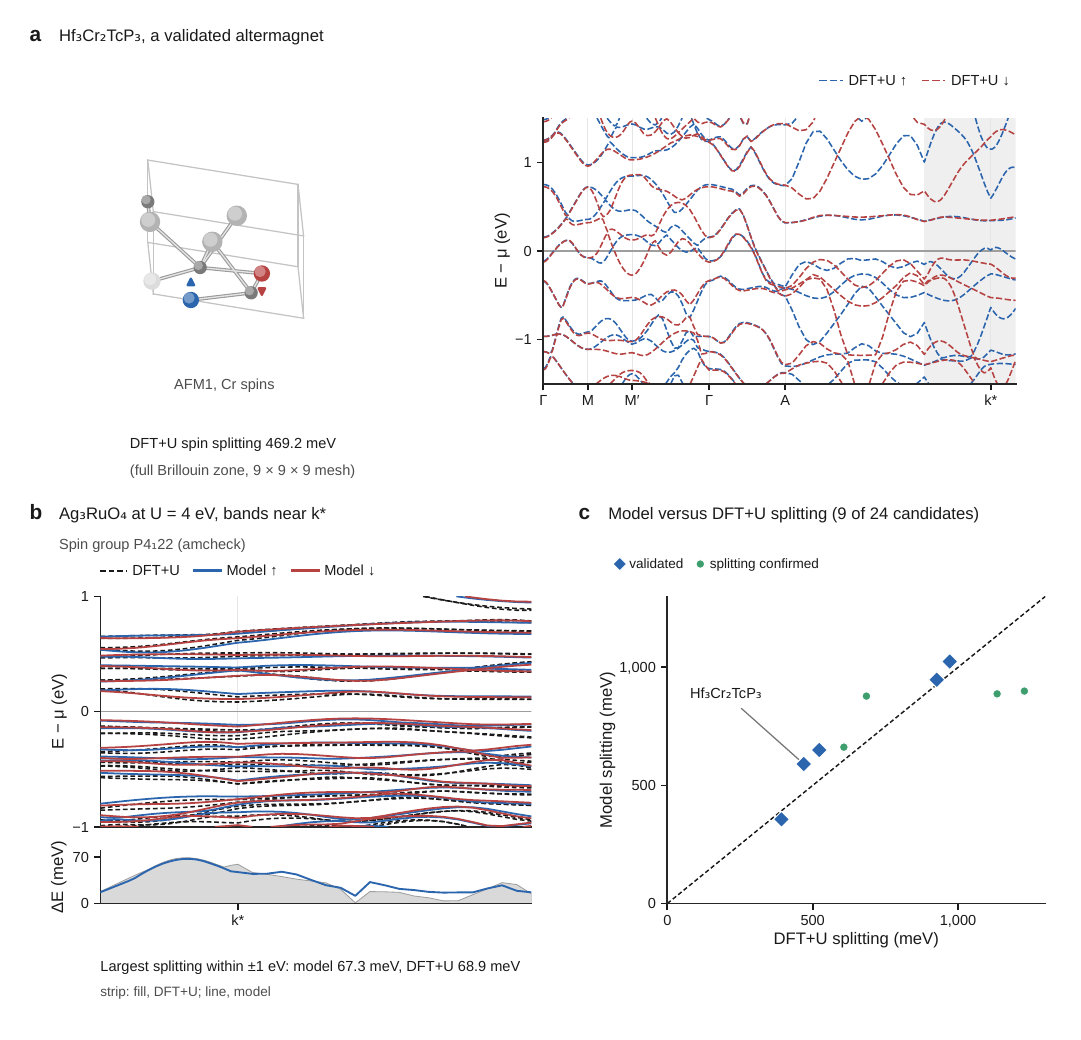}
\caption{\textbf{DFT+U validation of altermagnet candidates.} \textbf{a}, \ce{Hf3Cr2TcP3} (AFM1 spins on Cr), one of five validated altermagnets among the 24 GNoME altermagnet candidates with the highest predicted splitting, which were adjudicated by DFT+U under rules fixed before any result was seen (Supplementary Table S4); its DFT+U bands (dashed, as in b) with the tabulated U of each element (Methods), spin up (blue) and down (red), along Γ–M–M$'$–Γ–A of the hexagonal cell (labels as in Fig. 3b) and through k*. The DFT+U spin splitting over a 9 × 9 × 9 Brillouin-zone mesh (Methods) is 469.2 meV; the spin-group class is read on the transition-metal sublattice that carries the splitting. \textbf{b}, \ce{Ag3RuO4} at U = 4 eV around k*: DFT+U (dashed) and model bands, and the same-band splitting at each k (fill, DFT+U; line, model). Within 1 eV of the chemical potential the largest splitting is 68.9 meV in DFT+U and 67.3 meV in the model; spin group $P4_122$ (amcheck). \textbf{c}, Model against DFT+U spin splitting for the nine of the 24 candidates whose altermagnetic splitting DFT+U confirms, both measured on the same 9 × 9 × 9 mesh with the same rule (Methods): blue diamonds, the five validated altermagnets; green, the four candidates whose splitting is confirmed in the antiferromagnetic state, three of them ferrimagnetic with an uncompensated Nd 4f moment (ferromagnetic and supercell references in Supplementary Note 7). The other 15 candidates are listed in Supplementary Table S4.}
\label{fig:F6}
\end{figure}
\clearpage
\end{document}